\documentclass[a4,11pt]{article} 
\usepackage{amssymb}
\usepackage{caption}
\usepackage{longtable}
\usepackage{amsmath}
\usepackage[pdftex]{graphicx}
\usepackage{multirow}
\usepackage{cite}
\usepackage{slashed}
\usepackage{bm}
\usepackage{fancyhdr}
\usepackage{float}
\usepackage{tocloft}
\usepackage[top=2.5cm, bottom=2.0cm, left=2.5cm, right=2cm]{geometry}

\begin{document}

\begin{center}
{\large \bf	 Axion-like particle production at $\gamma e^{-}$ and $e^{-} e^{+} $ scatterings }\\

\vspace*{1cm}

 { Bui Thi Ha Giang$^{a, }$ \footnote{e-mail: giangbth@hnue.edu.vn, ORCID: 0000-0001-5814-0645} }\\

\vspace*{0.5cm}
 $^a$ Hanoi National University of Education, 136 Xuan Thuy, Hanoi, Vietnam
\end{center}

\begin{abstract}
This work explores the production and decay channels of axion-like particle (ALP) at upcoming high-energy accelerators, specifically the Future Circular Collider (FCC-ee),  International Linear Collider (ILC) and the Large Hadron Electron Collider (LHeC). Focusing on the benchmark parameter set $f_a = 1\text{ TeV}$, $C_{WW} = C_{BB} = 1$, $g_{aee} = 0.64$, and an ALP mass of $m_a = 10\text{ GeV}$, we calculate the differential and total cross-sections in $\gamma^{*}e^{-}$ and $e^{-}e^{+}$ collisions including the effect of the scattering angle, the polarization coefficients, the energy, the ALP mass and the scale $f_{a}$. From that, we compare the total cross-sections at the different accelerators. Our analysis highlights a strong complementarity among the proposed machines: the FCC-ee  ($\sqrt{s} = 91\text{ GeV}$) achieves the highest sensitivity in $e^-e^+ \to \gamma a$ with $\mathcal{O}(10^6)$ expected events via resonant enhancement, whereas the LHeC ($E_e = 60\text{ GeV}$) and the ILC ($\sqrt{s} = 1\text{ TeV}$) provide substantial signal yields of $\mathcal{O}(10^5)$ and $\mathcal{O}(10^4)$ events, respectively, through the $\gamma^* e^- \to a e^-$ subprocess. 
\end{abstract}
\textit{Keywords}: axion-like particle, cross-section, significance level.

\section{Introduction}
\hspace*{1cm} The discovery of Higgs boson at the LHC in 2012 seemingly completes the picture of elementary particle physics predicted by Standard model (SM). However, SM still presents some issues that require the existence of new physics. One of the drawback of SM is the so-called strong charge-parity (CP) problem \cite{jhep03, abel}. Peccei-Quinn (PQ) introduced a new global U(1) symmetry to resolve the CP problem \cite{pec1, pec2}. As a result of the U(1) symmetry breaking, the existence of axion is predicted. Given the energy scale for the symmetry breaking higher than that of the electroweak symmetry, the coupling of axion to ordinary matter can be feeble. Therefore, axion can be a natural candidate for cold dark matter (DM) \cite{kill, dine, abbott}. Recently, the axion has been considered in and the study about axion has extended to axion-like particles (ALPs). In addition to axion, more generic ALPs, denoted as $a$, are also predicted in various contexts. \\
\hspace*{1cm} ALPs, good DM candidates, are capable of producing detectable signatures across a broad range of experimental setups \cite{chigusa}. The evolution of stars and of core-collapse supernova can show the signal of the existence of ALPs. The interesting regions in the ALP parameter space has motivated ALP searches in the physics program of neutrino experiments \cite{sierra}. The coupling between ALPs and leptons to cosmology was studied particularly in the Big Bang Nucleosynthesis (BBN) \cite{ghosh}. The mass spectrum of ALPs and the associated energy scales of new physics span an extensive range from sub-eV values to the TeV domain corresponding to underlying dynamics that can originate near the electroweak scale or extend well beyond it. In the low-mass ALPs, particularly those in the eV to MeV region, have significant implications for early universe cosmology and stellar evolution. Their potential influence on phenomena such as the Cosmic Microwave Background (CMB), BBN, and stellar cooling has led to stringent constraints on their interactions with SM fields, derived from astrophysical and cosmological data \cite{mosana}. The ALP-photon coupling has been focused as a method of searching for ALPs in many experiments \cite{ehret, homma, osqar, ejlli, sapphires}. There are some models which have provided motivation to search for light ALPs, with masses in the range between an MeV and tens of GeV \cite{bauer3}. In the high mass domain, the MeV to TeV mass window, laboratory searches for ALPs have been conducted through beam dump experiments at CLEO and BaBar \cite{rich, riordan, ken, ball}. ALP signatures has been investigated at high energy colliders in the accelerators, beginning with LEP and continuing at the LHC in recent years \cite{opal, masu, atlas, mbauer1, mbauer2} as well as in the future experimental programs \cite{na62, gard, berlin, chou, curtin, aieli, na64}. In particular, the high intensity photon flux and associated electromagnetic cascades in reactor and accelerator neutrino experiments have motivated to explore ALP production including Primakoff scattering, Compton-like scattering, $e^{+}e^{-}$ annihilation, ALP bremsstrahlung \cite{ccm23}.  \\
\hspace*{1cm} In this work, we consider the colliders at the International Linear Collider (ILC), the Large Hadron electron Collider (LHeC), and the Future Circular Collider in its electron-positron configuration (FCC-ee). ILC is a proposed next generation electron-positron collider. The ILC aims to study the Standard Model (SM) at the center-of-mass energy $\sqrt{s}$ of 250 GeV, 500 GeV and possibly 1 TeV \cite{corsu}. The ILC, in addition to being an  electron-positron collider, can also operate effectively as a $\gamma\gamma$ collider, by virtue of the equivalent photon approximation. In parallel, LHeC has been  put forward as a complementary facility to the LHC \cite{fer, bruen, tini}. The clean experimental environment and the prospects for the substantial extension of the kinematic range in deep inelastic scattering enhance to find new physics beyond the SM (BSM). The ALP is clearly an interesting BSM scenario that is worthy of being studied at the LHeC \cite{xyue}. Moreover, FCC-ee is a proposed high luminosity circular lepton collider designed to operate over a broad center-of-mass energy range of approximately 88-365 GeV \cite{aba}. Owing to its well-defined initial state energy, high luminosity, and relatively low hadronic background compared with hadron colliders, FCC-ee provides a particularly favorable environment for precision measurements and searches for rare processes or weakly coupled new particles, including ALPs \cite{aba, bac}.\\
\hspace*{1cm} The structure of our work is the following: in section 2, we review about the framework of effective field theory (EFT) after electroweak (EW) symmetry breaking. In section 3, we investigate the cross-section in ALP production at $\gamma^{*} e^{-} \rightarrow ae^{-} $ and $e^{-}e^{+} \rightarrow \gamma a $ collisions. In section 4, we calculate the decay of ALP 10 GeV. Finally, section 5 closes our analysis with some concluding remarks.
\section{Framework}
ALPs are pseudo-scalars that feebly couple to SM particles \cite{sierra}. The ALP effective Lagrangian can be expressed as follows \cite{bieko}:
\begin{equation}
\begin{aligned}
\mathcal{L} = &\frac{1}{2}({\partial_\mu }a)({\partial^\mu }a) - \frac{{m_a^2}}{2}{a^2} + {C_{GG}}\frac{{{\alpha _S}}}{{4\pi {f_a}}}aG_{\mu \nu }^A{{\bar G}^{\mu \nu A}} + {C_{WW}}\frac{{{\alpha _L}}}{{4\pi {f_a}}}aW_{\mu \nu }^I{{\bar W}^{\mu \nu I}} + {C_{BB}}\frac{{{\alpha _Y}}}{{4\pi {f_a}}}a{B_{\mu \nu }}{{\bar B}^{\mu \nu }}\\
&+ \frac{{{\partial ^\mu }a}}{{{f_a}}}\sum\limits_F {\bar F} {C_F}{\gamma _\mu }F + \frac{{{C_H}}}{{{f_a}}}{\partial ^\mu }a({H^ + }i \overleftrightarrow{D}_{\mu} H) + \frac{{{C_{HH}}}}{{{f_a}^2}}({\partial ^\mu }a)({\partial _\mu }a){H^ + }H.
\end{aligned}
\end{equation}

In this framework, the matrices $C_{F} (F = q, u, d, l, e)$ are 3 x 3 Hermitian matrices, while the coefficients $C_{ii} (i = G, W, B, H)$ are real couplings. Factors of $\alpha_{i}/4\pi$ have pulled out from the interaction terms with the SM gauge fields. The sum extends over the chiral fermion multiplets F of the SM. It is worth noting that the couplings of ALP to bosonic fields are loop suppressed, while the fermionic ALP couplings are present at tree level.\\
After electroweak symmetry breaking, the ALP couplings to SM mass eigenstates are
\begin{equation}
\begin{aligned}
\mathcal{L} (\mu_{EW}) = & \dfrac{a}{f_{a}} \left(g_{agg} G_{\mu \nu }^A{\bar G}^{\mu \nu A} + g_{a\gamma\gamma} F_{\mu\nu} {\bar F}^{\mu \nu} + g_{a\gamma Z} F_{\mu\nu} {\bar Z}^{\mu \nu} + g_{aZZ} Z_{\mu\nu} {\bar Z}^{\mu \nu} + g_{aWW} W^{+}_{\mu\nu} {\bar W}^{\mu \nu} \right)\\
&+ \frac{2e}{s_{W}} \dfrac{a}{f_{a}} g_{aWW}\left(W_{\mu}^{+}W^{-}_{\nu} \left(s_{W}{\bar F}^{\mu \nu} + c_{W}{\bar Z}^{\mu \nu} \right) + \left( s_{W}A_{\mu} + c_{W} Z_{\mu} \right) \left(W_{\nu}^{+} {\bar W}^{\mu \nu}_{-} - W_{\nu}^{-} {\bar W}^{\mu \nu}_{+}\right)\right)  \\
& + \frac{1}{2}({\partial_\mu }a)({\partial^\mu }a) - \frac{{m_a^2}}{2}{a^2} + \dfrac{c_{HH}}{2f_{a}^{2}} ({\partial_\mu }a)({\partial^\mu }a) (h^{2} + 2\nu h) + \mathcal{L}_{ferm} (\mu_{EW}),
\end{aligned}
\end{equation}
where the electromagnetic coupling constant e, the sine and cosine of the Weinberg angle are denoted by $s_{W}$ and $c_{W}$, respectively. Due to the limit of infinite fermion masses, the couplings become constants as follows:
 \begin{align}
 &g_{agg} = \dfrac{\alpha_{s}}{4\pi} C_{GG},\\
 &g_{a\gamma\gamma} = \dfrac{\alpha}{4\pi} \left(C_{BB} + C_{WW}\right),\\
 &g_{aZZ} = \dfrac{\alpha}{4\pi s^{2}_{W}c^{2}_{W}} \left(c^{4}_{W}C_{WW} + s^{4}_{W}C_{BB}\right),\\
 &g_{aWW} = \dfrac{\alpha}{2\pi s^{2}_{W}} C_{WW},\\
 &g_{a\gamma Z} = \dfrac{\alpha}{2\pi s_{W}c_{W}} \left(c^{2}_{W}C_{WW} - s^{2}_{W}C_{BB}\right),
 \end{align}
where $C_{BB}$ and $C_{WW}$ are two Wilson coefficients.
\section{The cross-sections for ALP production in the high energy collisions}
\subsection{$\gamma^{*} e^{-} \rightarrow ae^{-} $ collision}
\hspace*{1cm} We consider the ALP production in the $\gamma^{*} e^{-}$ collision at the ILC and LHeC. The potential of the ILC can be further enhanced by $\gamma^{*} e^{-}$ scattering with the photon beam generated by the backward Compton scattering of incident electron and laser beam. While the LHeC can also be transformed to $\gamma^{*} e^{-}$ collision with the photon beam radiated from proton and the radiating proton remaining intact \cite{yue}. 
\begin{equation} \label{pt3.1}
 e^{-} (p_{1}) + \gamma^{*} (p_{2}) \rightarrow e^{-} (k_{1}) + a (k_{2}).
\end{equation}
\hspace*{1cm}The transition amplitude representing the s-channel is given by
\begin{equation}
M_{s} =  \varepsilon_{\mu} (p_{2})(-ie\gamma^{\mu}) u(p_{1})  \dfrac{i\left(\slashed{q}_{s} + m_{e}\right)}{q^{2}_{s} - m^{2}_{e}}\overline{u}(k_{1})g_{aee}\gamma^{5}.
\end{equation}
\hspace*{1cm}The transition amplitude representing the u-channel can be written as
\begin{equation}
M_{u} = u(p_{1})g_{aee}\gamma^{5}\dfrac{i\left(\slashed{q}_{u} + m_{e}\right)}{q^{2}_{u} - m^{2}_{e}}\varepsilon_{\mu}(p_{2})(-ie\gamma^{\mu})\overline{u}(k_{1}).
\end{equation}
\hspace*{1cm}The transition amplitude representing the t-channel is given by
\begin{equation}
M_{t} = M_{tZ} + M_{t\gamma},
\end{equation}
here
\begin{align}
&M_{tZ} = u(p_{1})(i\overline{g}_{eeZ})\gamma_{\nu} (-1 + 4s^{2}_{W} + \gamma^{5}) \overline{u}(k_{1})\dfrac{-i}{q^{2}_{t} - m^{2}_{Z}}\left(\eta^{\nu\sigma} - \dfrac{q_{t}^{\nu} q_{t}^{\sigma}}{m^{2}_{Z}}\right)\dfrac{2ig_{a\gamma Z}}{f_{a}}\varepsilon_{\mu\sigma\alpha\beta} p_{2}^{\alpha} q_{t}^{\beta}\varepsilon^{\mu}(p_{2}),\\
&M_{t\gamma} = u(p_{1})(-ie\gamma_{\nu})\overline{u}(k_{1})\dfrac{-i}{q^{2}_{t}}\eta^{\nu\sigma}\dfrac{4ig_{a\gamma\gamma}}{f_{a}}\varepsilon_{\mu\sigma\alpha\beta} p_{2}^{\alpha} q_{t}^{\beta}\varepsilon^{\mu}(p_{2}).
\end{align}
\hspace*{1cm} The effective cross-section $\sigma(s)$ for the $ \gamma^{*} e^{-} \rightarrow ae^{-}$ sub-process at the ILC can be calculated as follows \cite{yue}
\begin{equation}
\sigma (s) = \int_{(m_{a} + m_{e})^{2}/s}^{0.83} dx_{s} f_{\gamma/e}(x_{s}) \int_{(cos\theta)_{min}}^{(cos\theta)_{max}} d cos\theta \dfrac{d\widehat{\sigma}(\widehat{s})}{d cos\theta},
\end{equation} 
where 
\begin{equation}
\frac{d{\widehat{\sigma}(\widehat{s})}}{d(cos\theta)} = \frac{1}{32 \pi \widehat{s}} \frac{|\overrightarrow{k}_{1}|}{|\overrightarrow{p}_{1}|} |M_{fi}|^{2}
\end{equation}
is the expressions of the differential cross-section, $x_{s} = \widehat{s}/s$ in which $\sqrt{\widehat{s}}$ is center-of-mass energy of the $\gamma^{*}e^{-} \rightarrow ae^{-}$ sub-process, $\sqrt{s}$ is center-of-mass energy of the ILC experiments, $\theta = (\widehat{\overrightarrow{p}_{1}, \overrightarrow{k}_{2}})$ is the scattering angle, $x_{s max} = \dfrac{\zeta}{1 + \zeta}$. The photon distribution function $f_{\gamma/e}$ is given by \cite{ginz}
\begin{equation}
f_{\gamma/e} = \dfrac{1}{D(\zeta)}\left[(1 - x_{s})+\dfrac{1}{1 - x_{s}} - \dfrac{4x_{s}}{\zeta(1 - x_{s})} + \dfrac{4x_{s}^{2}}{\zeta^{2}(1 - x_{s})^{2}} \right],
\end{equation}
where 
\begin{equation}
D(\zeta) = \left(1 - \dfrac{4}{\zeta} - \dfrac{8}{\zeta^{2}}\right)ln(1 + \zeta) + \dfrac{1}{2} + \dfrac{8}{\zeta} - \dfrac{1}{2(1 + \zeta)^{2}}.
\end{equation}
For $\zeta = 4.8$, $x_{s max} = 0.83$. The scattering angle is chosen as $10^{0} \leq \theta \leq 170^{0}$ to make the scattered particles be detected \cite{yue}. We give some estimates for the cross-sections at ILC as follows:\\
\begin{figure}[!htb] 
\begin{center}
       \begin{tabular}{cc}
        \includegraphics[width=6.5cm, height= 5cm]{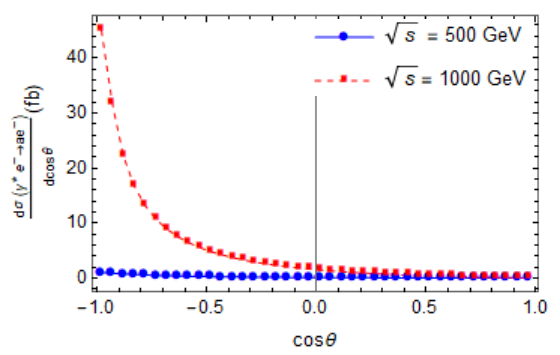} &
        \includegraphics[width=6.5cm, height= 5cm]{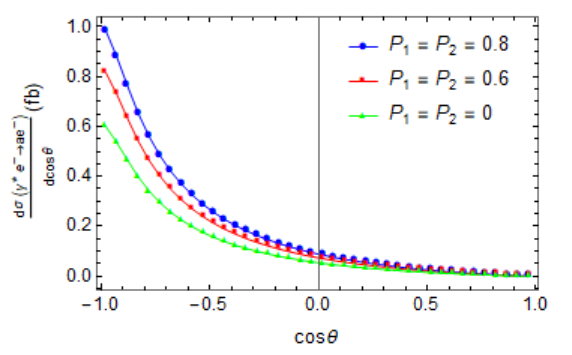} \\
        (a) & (b)
    \end{tabular} 
             \caption{\label{Fig.1} The differential cross-section as a function of the $cos\theta$ in $\gamma^{*} e^{-} \rightarrow ae^{-} $ collision at ILC.}
\end{center}
\end{figure}
\hspace*{0.5cm} Initially, in Fig.\ref{Fig.1}, we evaluate the dependence of the differential cross-sections on the $cos\theta$. The parameters are chosen as $f_{a} = 1$ TeV, $C_{WW} = C_{BB} = 1$ \cite{bieko}, $g_{aee} = 0.64$, the ALP mass $m_{a} = 10$ GeV \cite{adh}. The polarization coefficients of initial and final electron beams in Fig.\ref{Fig.1}a are $P_{1} = P_{2} = 0.8$ \cite{brit}. When the center-of-mass energy is chosen as $\sqrt{s} = 1000$ GeV (Fig.\ref{Fig.1}a), the differential cross-section decreases fast when $-1 \lesssim cos\theta \lesssim -0.5$, then decreases gradually when $-0.5 \lesssim cos\theta \lesssim 1$. In Fig.\ref{Fig.1}b the polarization coefficients of initial and final electron beams are $(P_{1}, P_{2}) = (0.8, 0.8); (0.6, 0.6); (0, 0)$, respectively, the center-of-mass energy is chosen as $\sqrt{s} = 500$ GeV. The differential cross-sections decrease gradually when $-1 \lesssim cos\theta \lesssim 1$ in Fig.\ref{Fig.1}b. Therefore, the differential cross-sections are peaked in the backward direction, but flat in the forward direction, which is advantageous to collect ALP signals from experiment.\\
\begin{figure}[!htb] 
\begin{center}
       \begin{tabular}{cc}
        \includegraphics[width=6.5cm, height= 5cm]{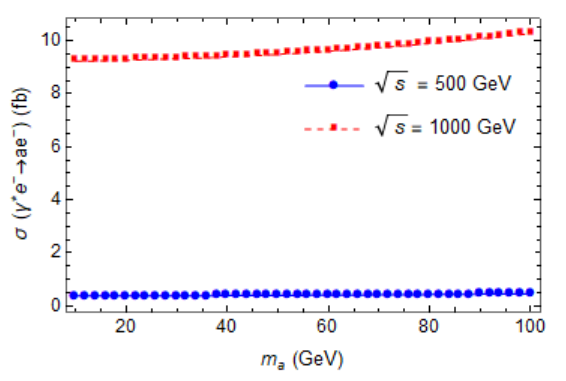} &
        \includegraphics[width=6.5cm, height= 5cm]{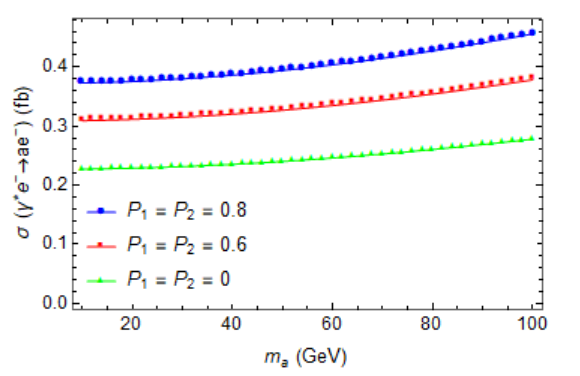} \\
        (a) & (b)
    \end{tabular} 
             \caption{\label{Fig.2} The total cross-section as a function of the ALP mass $m_{a}$ in $\gamma^{*} e^{-} \rightarrow ae^{-} $ collision at ILC.}
\end{center}
\end{figure}
\hspace*{0.5cm} Secondly, in Fig.\ref{Fig.2}, we plot the dependence of total cross-sections on the ALP mass. The parameters in Fig.\ref{Fig.2} are chosen as Fig.\ref{Fig.1}. A marginal increase in the total cross-sections is observed as the ALP mass from 10 GeV to 100 GeV.\\
\begin{figure}[!htb] 
\begin{center}
       \begin{tabular}{cc}
        \includegraphics[width=6.5cm, height= 5cm]{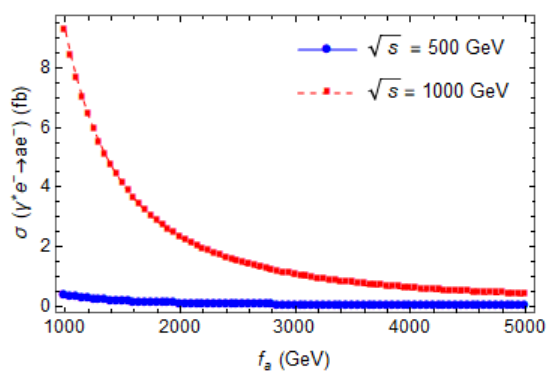} &
        \includegraphics[width=6.5cm, height= 5cm]{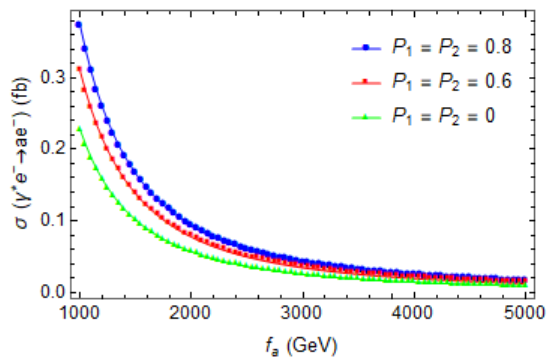} \\
        (a) & (b)
    \end{tabular} 
             \caption{\label{Fig.3} The total cross-section as a function of the $f_{a}$ in $\gamma^{*} e^{-} \rightarrow ae^{-} $ collision at ILC.}
\end{center}
\end{figure}
\hspace*{0.5cm} Along the same lines, the total cross-sections are evaluated as the function of the scale $f_{a}$ in Fig.\ref{Fig.3} with the parameters chosen as Fig.\ref{Fig.1}. As shown in Fig.\ref{Fig.3}a, the total cross-section decreases rapidly with $f_a$ ranging from 1000 to 5000 GeV at $\sqrt{s} = 1000$ GeV. In contrast, for $\sqrt{s} = 500$ GeV and various polarization coefficients (Fig.\ref{Fig.3}b), the cross-sections exhibit a more gradual decline over the same $f_a$ range.\\

\begin{figure}[!htb] 
\begin{center}
               \includegraphics[width=6.5 cm, height= 5cm]{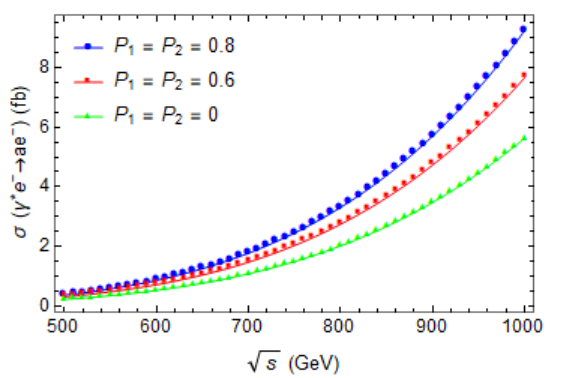} 
                         \caption{\label{Fig.4} The total cross-section as a function of the center-of-mass energy $\sqrt{s}$ in $\gamma^{*} e^{-} \rightarrow ae^{-} $ collision at ILC. }
\end{center}
\end{figure}
\hspace*{0.5cm} Next, the parameters in Fig.\ref{Fig.4} are chosen as Fig.\ref{Fig.1}, and the polarization coefficients of initial and final electron beams are $(P_{1}, P_{2}) = (0.8, 0.8); (0.6, 0.6); (0, 0)$, respectively. Fig.\ref{Fig.4} indicates that the total cross-sections increase when the center-of-mass energy $\sqrt{s}$ increases in range of $500  \text{GeV} \leq \sqrt{s} \leq 1000  \text{GeV}$.\\
\begin{figure}[!htb] 
\begin{center}
              \includegraphics[width=7.5cm, height= 5cm]{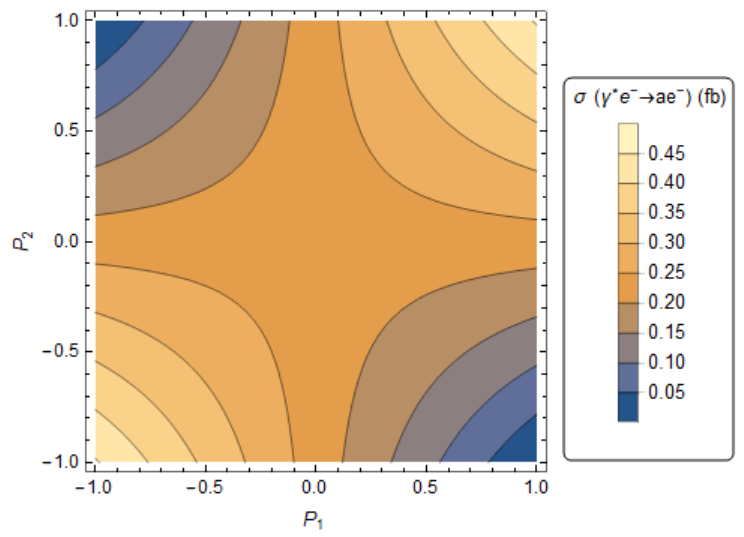}     
             \caption{\label{Fig.5} The total cross-section as a function of the polarization coefficients of initial and final electron beams in $\gamma^{*} e^{-} \rightarrow ae^{-} $ collision at ILC.}
\end{center}
\end{figure}
\hspace*{0.5cm} Lastly, the total cross-section is evaluated as the function of the polarization coefficients of initial and final electron beams at the ILC in Fig.\ref{Fig.5}. The parameters are chosen as $f_{a} = 1$ TeV, $C_{WW} = C_{BB} = 1$ \cite{bieko}, $g_{aee} = 0.64$, the ALP mass $m_{a} = 10$ GeV \cite{adh}, the center-of-mass energy $\sqrt{s} = 500$ GeV. The total cross-section reaches its maximum value (represented by the yellow region) when $P_{1} = P_{2} = \pm 1$. Conversely, it attains its minimum (indicated by the dark blue region) when $P_{1} = 1, P_{2} = -1$ and vice versa.\\

\hspace*{1cm} To continue, in the next part, we evaluate the cross-section at the LHeC 14 TeV. The effective production cross-section for the sub-process at the LHeC can be described \cite{yue}
\begin{equation}
\sigma (\gamma^{*} e^{-} \rightarrow  a e^{-}) = \int^{\xi_{max}}_{Max(\dfrac{(m_{e} + m_{a})^{2}}{s}, \xi_{min)}}E_{p}f(\xi E_{p})d\xi\int^{(cos\theta)_{max}}_{(cos\theta)_{min}}\dfrac{d\widehat{\sigma}(\widehat{s})}{dcos\theta}dcos\theta,
\end{equation} 
where $\widehat{s} = 4E_{e}E_{\gamma} = \xi s$ is center of mass energy of the sub-process $\gamma^{*} e^{-} \rightarrow  a e^{-}$. The photon flux can be written as
\begin{equation}
f(\xi E_{p}) = \int_{Q^{2}_{min}}^{Q^{2}_{max}}\dfrac{dN_{\gamma}}{dE_{\gamma}dQ^{2}}dQ^{2},
\end{equation}
where
\begin{equation}
\dfrac{dN_{\gamma}}{dE_{\gamma}dQ^{2}} = \dfrac{\alpha_{e}}{\pi}\dfrac{1}{E_{\gamma}Q^{2}}\left[\left(1-\dfrac{E_{\gamma}}{E_{p}}\right)\left(1-\dfrac{Q^{2}_{min}}{Q^{2}} \right)F_{E}+\dfrac{E^{2}_{\gamma}}{2E^{2}_{p}}F_{M}\right],
\end{equation}
with 
\begin{align}
&Q^{2}_{min} = \dfrac{M^{2}_{P}E^{2}_{\gamma}}{E_{p}(E_{p}-E_{\gamma})},\\
&F_{E} = \dfrac{4M^{2}_{P}G^{2}_{E} + Q^{2}G_{M}^{2}}{4M^{2}_{P} + Q^{2}},\\
&G^{2}_{E} = \dfrac{G^{2}_{M}}{\mu^{2}_{P}} = \left(1+\dfrac{Q^{2}}{Q^{2}_{0}} \right)^{-4},\\
&F_{M} =  G^{2}_{M}.
\end{align}
$M_{P}$ is the mass of the proton. $E_{p}$ is the energy of the incoming proton beam. $E_{\gamma} = \xi E_{p}$ is the photon energy, which is related to the loss energy of the emitted proton beam. $F_{E}, F_{M}$ are functions of the electric and magnetic form factors given in the dipole approximation. $\mu^{2}_{P} = 7.78$ is the magnetic moment of the proton. \\
\hspace*{1cm}For numerical calculations, we choose the energy of the incoming proton beam at LHeC $E_{p} = 7$ TeV, the electron beam energy $E_{e}$ in the range 60 GeV $ \leq E_{e} \leq 140$ GeV \cite{rod}. Since the contribution to the above integral formula is very small for $Q^{2}_{max} > 2 GeV^{2}$, therefore $Q^{2}_{max}$ is chosen as 2 $GeV^{2}$ \cite{yue}. We give estimates for the cross-sections as follows \\
\begin{figure}[!htb] 
\begin{center}
       \begin{tabular}{cc}
        \includegraphics[width=7cm, height= 4.5cm]{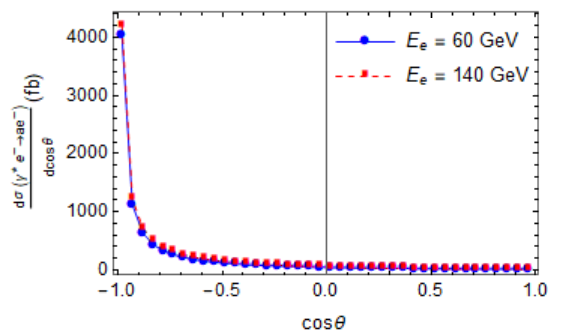} &
        \includegraphics[width=7cm, height= 4.5cm]{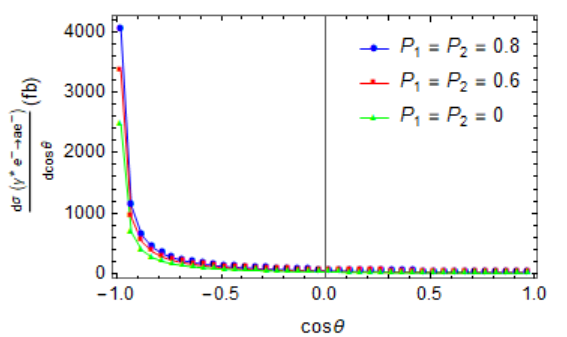} \\
        (a) & (b)
    \end{tabular} 
             \caption{\label{Fig.6} The differential cross-section as a function of the $cos\theta$ in $\gamma^{*} e^{-} \rightarrow ae^{-} $ collision at LHeC.}
\end{center}
\end{figure}
\hspace*{0.5cm} Firstly, in Fig.\ref{Fig.6}, we evaluate the dependence of the differential cross-section on the $cos\theta$. The parameters are chosen as $f_{a} = 1$ TeV, $C_{WW} = C_{BB} = 1$ \cite{bieko}, $g_{aee} = 0.64$, the ALP mass $m_{a} = 10$ GeV \cite{adh}. While the polarization coefficients of initial and final electron beams are $P_{1} = P_{2} = 0.8$ in Fig.\ref{Fig.6}a, the electron beam energy is chosen as $E_{e} = 60$ GeV in Fig.\ref{Fig.6}b. The differential cross-section decreases fast when $-1 \lesssim cos\theta \lesssim -0.8$, then decreases gradually when $-0.8 \lesssim cos\theta \lesssim 1$. Therefore, the differential cross-sections are peaked in the backward direction, but flat in the forward direction, which is advantageous to collect ALP signals from experiment.\\

\begin{figure}[!htb] 
\begin{center}
       \begin{tabular}{cc}
        \includegraphics[width=6.5cm, height= 4.5cm]{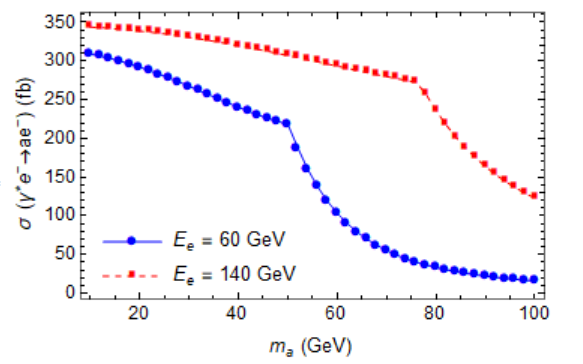} &
        \includegraphics[width=6.5cm, height= 4.5cm]{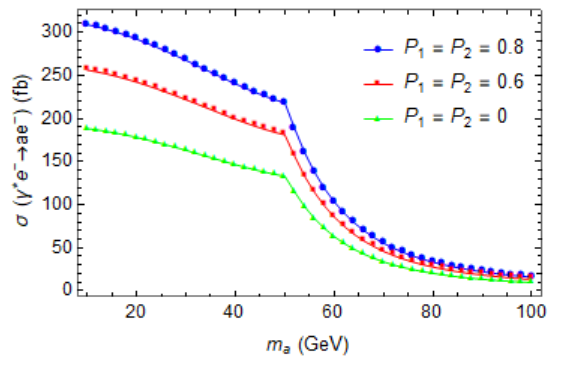} \\
        (a) & (b)
    \end{tabular} 
             \caption{\label{Fig.7} The total cross-section as a function of the ALP mass $m_{a}$ in $\gamma^{*} e^{-} \rightarrow ae^{-} $ collision at LHeC.}
\end{center}
\end{figure}
\hspace*{0.5cm} Similar to Fig.\ref{Fig.2}, we plot the dependence of total cross-sections on the ALP mass at LHeC in Fig.\ref{Fig.7}. The parameters are chosen as Fig.\ref{Fig.6}. While the polarization coefficients of initial and final electron beams are $P_{1} = P_{2} = 0.8$ in Fig.\ref{Fig.7}a, the electron beam energy is chosen as $E_{e} = 60$ GeV in Fig.\ref{Fig.7}b. At electron beam energy of 60 GeV, the total cross-section decreases gradually for $m_{a}$ between 10 GeV and approximately 50 GeV, followed by a rapid decline as $m_a$ increases toward 100 GeV. In contrast, at electron beam energy of 140 GeV, the total cross-section exhibits a moderate decrease in the range $10 \leq m_{a} \preceq 78$ GeV, before dropping sharply between 78 GeV and 100 GeV.\\

\begin{figure}[!htb] 
\begin{center}
       \begin{tabular}{cc}
        \includegraphics[width=6.5cm, height= 4.5cm]{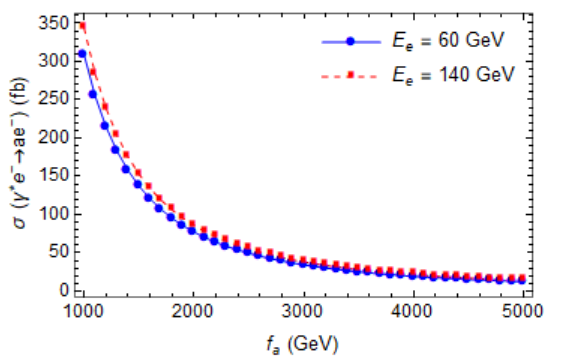} &
        \includegraphics[width=6.5cm, height= 4.5cm]{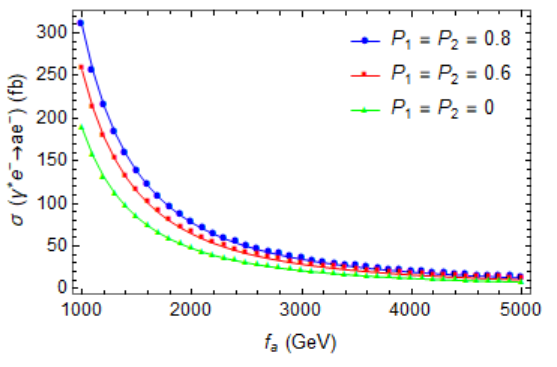} \\
        (a) & (b)
    \end{tabular} 
             \caption{\label{Fig.8} The total cross-section as a function of the $f_{a}$ in $\gamma^{*} e^{-} \rightarrow ae^{-} $ collision at LHeC.}
\end{center}
\end{figure}
\hspace*{0.5cm} Along the same lines, the total cross-sections are evaluated as the function of the scale $f_{a}$ in Fig.\ref{Fig.8} with the parameters chosen as Fig.\ref{Fig.6}. While the polarization coefficients of initial and final electron beams are $P_{1} = P_{2} = 0.8$ in Fig.\ref{Fig.8}a, the electron beam energy is chosen as $E_{e} = 60$ GeV in Fig.\ref{Fig.8}b. As shown in Fig.\ref{Fig.8}, the total cross-section decreases with $f_a$ ranging from 1000 to 5000 GeV.\\

\begin{figure}[!htb] 
\begin{center}
             \includegraphics[width=7cm, height= 5cm]{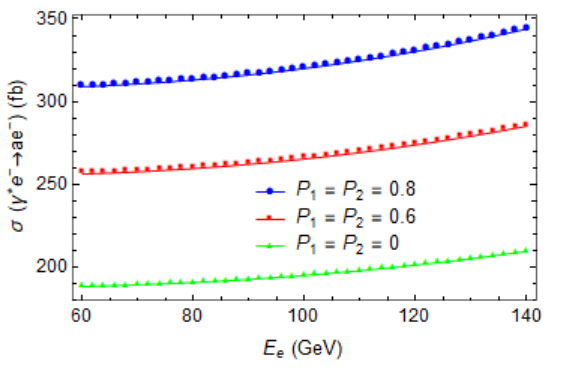} 
             \caption{\label{Fig.9} The total cross-section as a function of the electron beam energy $E_{e}$ in $\gamma^{*} e^{-} \rightarrow ae^{-} $ collision at LHeC. }
\end{center}
\end{figure}
\hspace*{0.5cm} In addition, the parameters in Fig.\ref{Fig.9} are chosen as Fig.\ref{Fig.6}, and the polarization coefficients of initial and final electron beams are $(P_{1}, P_{2}) = (0.8, 0.8); (0.6, 0.6); (0, 0)$, respectively. Fig.\ref{Fig.9} indicates that the total cross-sections increase gradually when the electron beam energy $E_{e}$ increases in range of $60 \text{GeV} \leq E_{e} \leq 140 \text{GeV}$.\\

\begin{figure}[!htb] 
\begin{center}
               \includegraphics[width=9cm, height= 6.5cm]{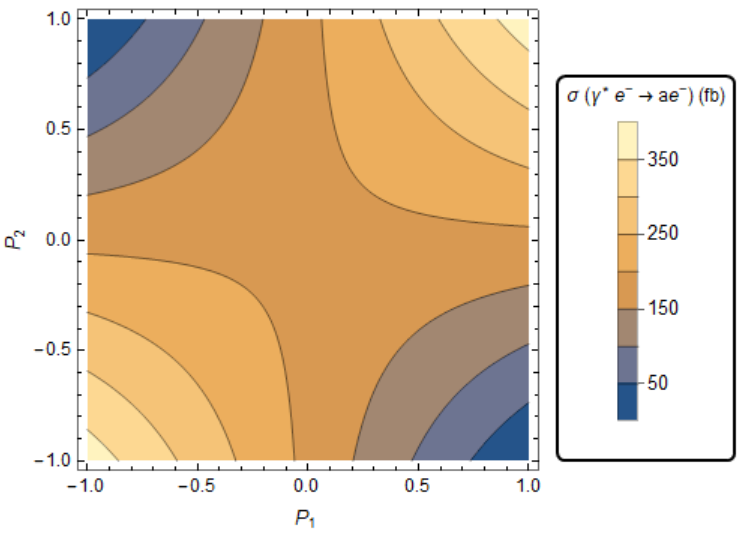} 
                    \caption{\label{Fig.10} The total cross-section as a function of the polarization coefficients of initial and final electron beams in $\gamma^{*} e^{-} \rightarrow ae^{-} $ collision at LHeC.}
\end{center}
\end{figure}
\hspace*{0.5cm} Finally, the total cross-section is evaluated as the function of the polarization coefficients of initial and final electron beams at the LHeC in Fig.\ref{Fig.10}. The parameters are chosen as $f_{a} = 1$ TeV, $C_{WW} = C_{BB} = 1$, $g_{aee} = 0.64$, the ALP mass $m_{a} = 10$ GeV, the electron beam energy $E_{e} = 60$ GeV. The total cross-section reaches its maximum value (represented by the yellow region) when $P_{1} = P_{2} = \pm 1$. Conversely, it attains its minimum (indicated by the dark blue) when $P_{1} = 1, P_{2} = -1$ and vice versa.\\

\subsection{$e^{-}e^{+} \rightarrow \gamma a $ collision}
\hspace*{1cm}We consider the scattering in which the initial state contains electron and positron,
the final state contains the photon and ALP.  
\begin{equation} \label{pt3.2}
 e^{-} (p_{1}) + e^{+} (p_{2}) \rightarrow \gamma (k_{1}) + a (k_{2}).
\end{equation}
\hspace*{1cm}The transition amplitude representing the s-channel is given by
\begin{equation}
M_{s} = M_{sZ} + M_{s\gamma},
\end{equation}
here
\begin{align}
&M_{sZ} = u(p_{1})\gamma_{\mu} (-1 + 4s^{2}_{W} + \gamma^{5}) (i\overline{g}_{eeZ}) \overline{v}(p_{2})\dfrac{-i}{q^{2}_{s} - m^{2}_{Z}}\left(\eta^{\mu\sigma} - \dfrac{q_{s}^{\mu}q_{s}^{\sigma}}{m_{Z}^{2}}\right)\varepsilon^{*\nu}(k_{1}) 2i \dfrac{g_{a\gamma Z}}{f_{a}} \varepsilon_{\sigma\nu\alpha\beta}q_{s}^{\alpha}k_{1}^{\beta},\\
&M_{s\gamma} = u(p_{1})(-ie\gamma_{\mu})\overline{v}(p_{2})\dfrac{-i}{q^{2}_{s}}\eta^{\mu\sigma}\varepsilon^{*\nu}(k_{1}) 4i  \dfrac{g_{a\gamma \gamma}}{f_{a}} \varepsilon_{\sigma\nu\alpha\beta}q_{s}^{\alpha}k_{1}^{\beta}.
\end{align}
\hspace*{1cm}The transition amplitude representing the u-channel can be written as
\begin{equation}
M_{u} = u(p_{1})g_{aee}\gamma^{5} \dfrac{i\left(\slashed{q}_{u} + m_{e}\right)}{q^{2}_{u} - m^{2}_{e}}\varepsilon^{*\nu}(k_{1})(-ie\gamma_{\nu})\overline{v}(p_{2}). 
\end{equation}
\hspace*{1cm}The transition amplitude representing the t-channel is given by
\begin{equation}
M_{t} = \overline{v}(p_{2}) g_{aee}\gamma^{5} \dfrac{i\left(\slashed{q}_{t} + m_{e}\right)}{q^{2}_{t} - m^{2}_{e}}\varepsilon^{*\nu}(k_{1})(-ie\gamma_{\nu})u(p_{1}) .
\end{equation}
\hspace*{1cm}
To compare the results obtained from the ILC and FCC-ee colliders, we further investigate the dependence of both differential and total cross-sections on the relevant parameters. We have detailed estimates for the cross-sections in $e^{-}e^{+} \rightarrow \gamma a$ collision at ILC and FCC-ee as follows: \\
\begin{figure}[!htb] 
\begin{center}
       \begin{tabular}{cc}
        \includegraphics[width=7.5cm, height= 5cm]{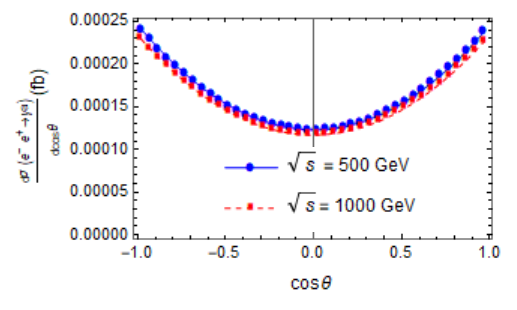} &
        \includegraphics[width=7.5cm, height= 5cm]{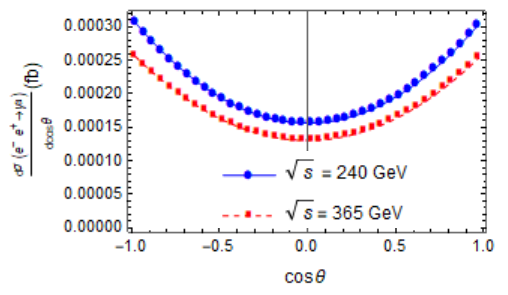} \\
        (a) & (b)\\
        \includegraphics[width=7.5cm, height= 5cm]{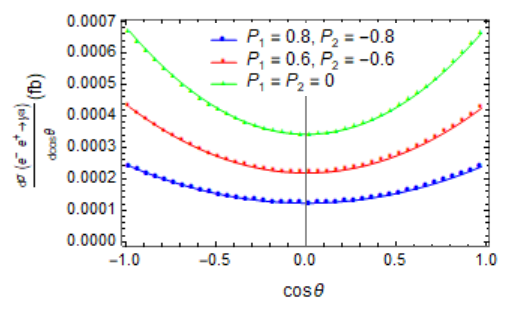} &
        \includegraphics[width=7.5cm, height= 5cm]{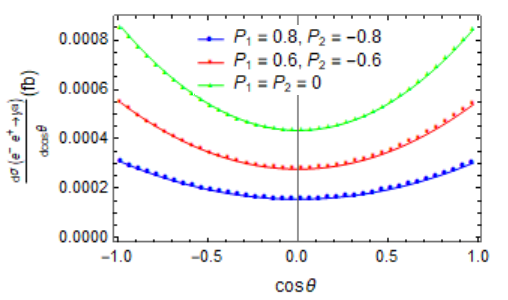} \\
        (c) & (d)        
    \end{tabular} 
             \caption{\label{Fig.11} The differential cross-section as a function of the $cos\theta$ in $e^{-}e^{+} \rightarrow \gamma a$ collision at ILC (a, c) and FCC-ee (b, d). }
\end{center}
\end{figure}
\hspace*{0.5cm} Looking at Fig.\ref{Fig.11}, we evaluate the dependence of the differential cross-section on the $cos\theta$. The parameters are chosen as $f_{a} = 1$ TeV, $C_{WW} = C_{BB} = 1$ \cite{bieko}, $g_{aee} = 0.64$, the ALP mass $m_{a} = 10$ GeV \cite{adh}. The polarization coefficients of initial electron and positron beams in Fig.\ref{Fig.11}a, b are $P_{e^-} = 0.8, P_{e^+} = -0.8$, respectively. The center-of-mass energy is chosen as $\sqrt{s} = 500$ GeV at the ILC, 240 GeV at the FCC-ee in Fig.\ref{Fig.11}c, d, respectively. The distribution shows that the differential cross-sections monotonically decrease in the backward region as $\cos\theta$ approaches zero from $-1$. After reaching a minimum at approximately $\cos\theta = 0$, the value monotonically increases in the forward region toward $\cos\theta = 1$.\\

\begin{figure}[!htb] 
\begin{center}
       \begin{tabular}{cc}
        \includegraphics[width=6.5cm, height= 4.5cm]{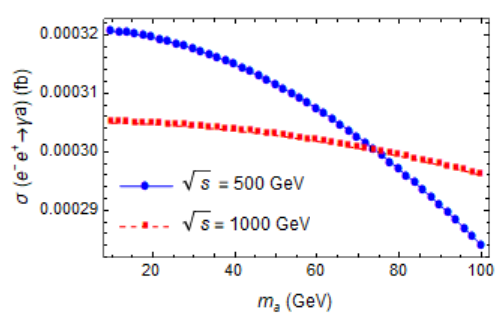} &
        \includegraphics[width=6.5cm, height= 4.5cm]{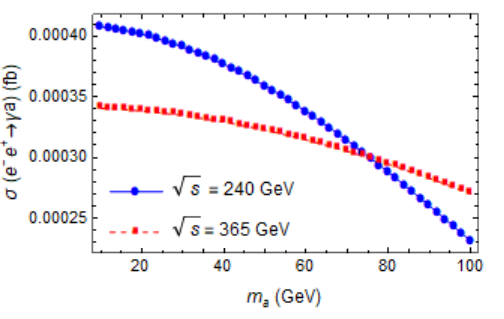} \\
        (a) & (b)\\
         \includegraphics[width=6.5cm, height= 4.5cm]{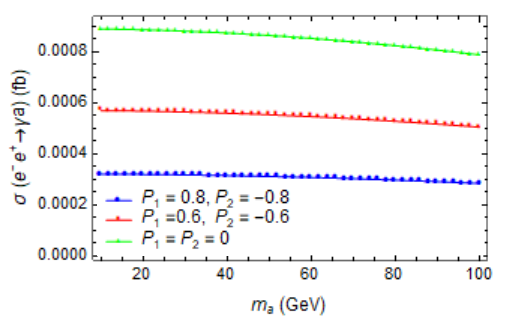} &
        \includegraphics[width=6.5cm, height= 4.5cm]{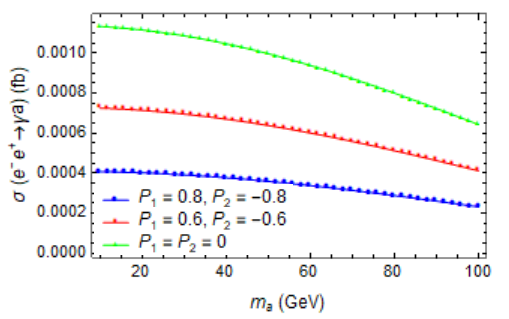} \\
        (c) & (d)
    \end{tabular}
             \caption{\label{Fig.12} The total cross-section as a function of the ALP mass $m_{a}$ in $e^{-}e^{+} \rightarrow \gamma a$ collision at ILC (a, c) and FCC-ee (b, d).}
\end{center}
\end{figure}
\hspace*{0.5cm} Moving to the evaluation of the total cross-sections, we plot the dependence of total cross-sections on the ALP mass in Fig.\ref{Fig.12}. The parameters in Fig.\ref{Fig.12} are chosen as Fig.\ref{Fig.11}. Overall, the total cross-sections decrease as the ALP mass $m_a$ increases from 10 GeV to 100 GeV. A similar trend is observed for both colliders: the cross-section exhibits a sharper decline at lower energies, specifically at $\sqrt{s} = 500$ GeV for the ILC (Fig.\ref{Fig.12}a) and $\sqrt{s} = 240$ GeV for the FCC-ee (Fig.\ref{Fig.12}b) whereas a more gradual reduction occurs at their respective higher energy stages.\\
\begin{figure}[!htb] 
\begin{center}
       \begin{tabular}{cc}
        \includegraphics[width=6.5cm, height= 4.5cm]{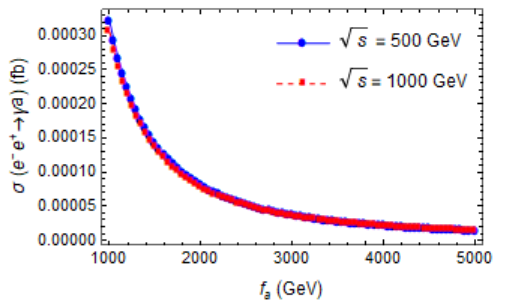} &
        \includegraphics[width=6.5cm, height= 4.5cm]{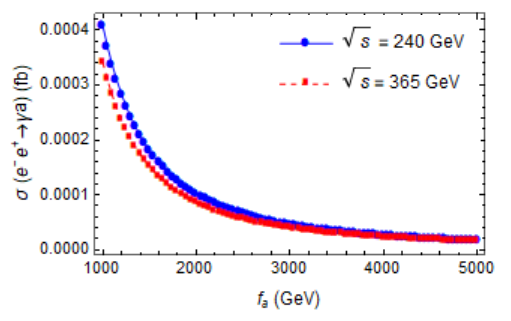} \\
        (a) & (b)\\
         \includegraphics[width=6.5cm, height= 4.5cm]{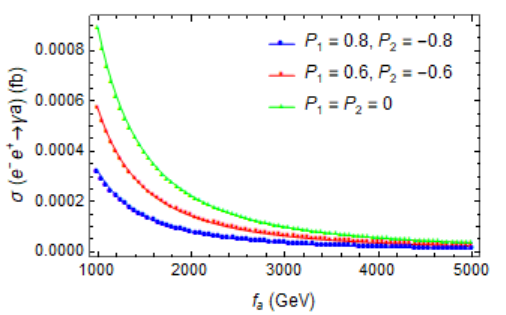} &
        \includegraphics[width=6.5cm, height= 4.5cm]{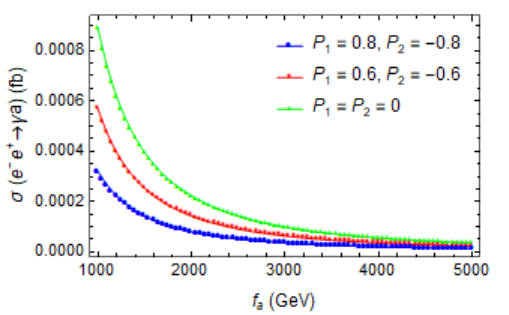} \\
        (c) & (d)
    \end{tabular}
             \caption{\label{Fig.13} The total cross-section as a function of the $f_{a}$ in $e^{-}e^{+} \rightarrow \gamma a$ collision at ILC (a, c) and FCC-ee (b, d).}
\end{center}
\end{figure}
\hspace*{0.5cm} In addition, the total cross-sections are evaluated as the function of the scale $f_{a}$ in Fig.\ref{Fig.13} with the parameters chosen as Fig.\ref{Fig.11}. As shown in Fig.\ref{Fig.13}, the total cross-sections are evaluated to decrease rapidly for $f_{a}$ values between 1000 and approximately 2500 GeV, followed by a more gradual decline as $f_{a}$ ranges from 2500 GeV to 5000 GeV.\\

\begin{figure}[!htb] 
\begin{center}
 \begin{tabular}{cc}
              \includegraphics[width=7cm, height= 5cm]{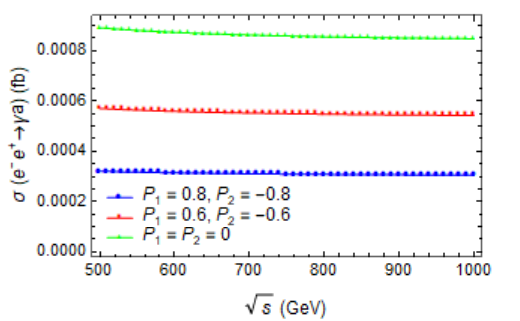} &
               \includegraphics[width=6.8cm, height= 5cm]{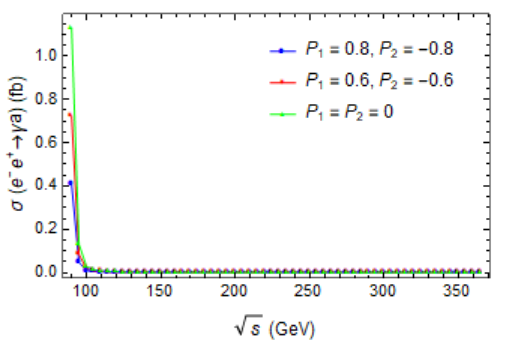} \\
               (a) & (b)
               \end{tabular}
                 \caption{\label{Fig.14} The total cross-section as a function of the center-of-mass energy $\sqrt{s}$ in $e^{-}e^{+} \rightarrow \gamma a$ collision at ILC (a) and FCC-ee (b).}
\end{center}
\end{figure}
\hspace*{0.5cm} Next, the parameters in Fig.\ref{Fig.14} are chosen as Fig.\ref{Fig.11}, and the polarization coefficients of initial electron and positron beams are $(P_{e^-}, P_{e^+}) = (0.8, -0.8); (0.6, -0.6); (0, 0)$, respectively. As illustrated in Fig.\ref{Fig.14}a, the reduction in the total cross-section tapers off as the energy increases from 500 to 1000 GeV at the ILC. In contrast, at the FCC-ee, the cross-section decreases sharply within the 10–100 GeV range and then declines more gradually as the energy scales from 100 to 365 GeV in Fig.\ref{Fig.14}b.\\
\begin{figure}[!htb] 
\begin{center}
\begin{tabular}{cc}
               \includegraphics[width=6.5cm, height= 5cm]{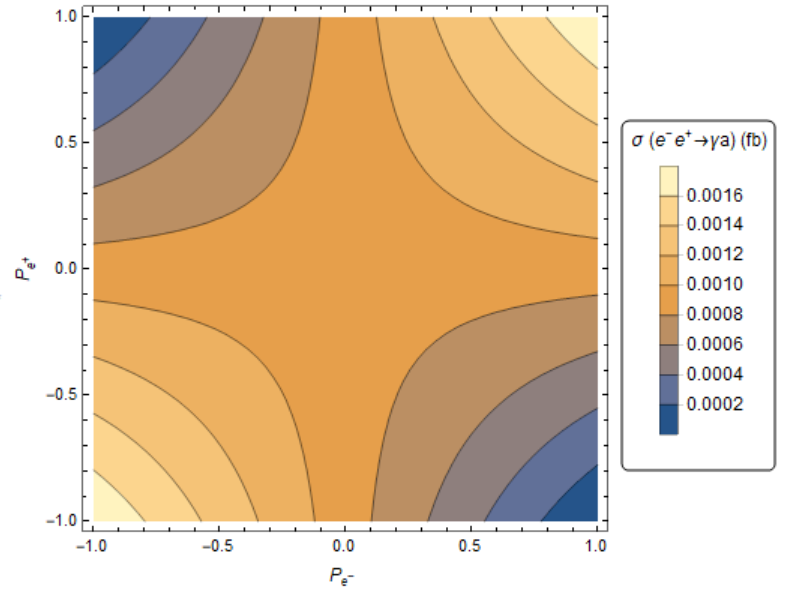} &
               \includegraphics[width=6.5cm, height= 5cm]{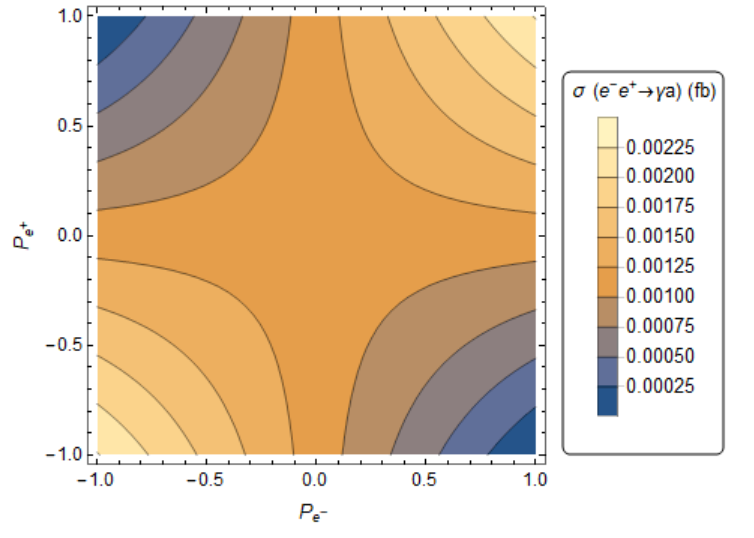} \\
               (a) & (b)
               \end{tabular}
             \caption{\label{Fig.15} The total cross-section as a function of the polarization coefficients of electron and positron beams in $e^{-}e^{+} \rightarrow \gamma a$ collision at ILC (a) and FCC-ee (b). }
\end{center}
\end{figure}
\hspace*{0.5cm} The total cross-section is evaluated as the function of the polarization coefficients of initial electron and positron beams at the ILC in Fig.\ref{Fig.15}a, and the FCC-ee in Fig.\ref{Fig.15}b, respectively. The parameters are chosen as $f_{a} = 1$ TeV, $C_{WW} = C_{BB} = 1$, $g_{aee} = 0.64$, the ALP mass $m_{a} = 10$ GeV. The center-of-mass energy is chosen as $\sqrt{s} = 500$ GeV (ILC) in Fig.\ref{Fig.15}a and $\sqrt{s} = 240$ GeV (FCC-ee) in Fig.\ref{Fig.15}b, respectively. The total cross-section reaches its maximum value (represented by the yellow regions) when $P_{e^-} = P_{e^+} = \pm 1$. Conversely, it attains its minimum (indicated by the dark blue regions) when $P_{e^-} = 1, P_{e^+} = -1$ and vice versa.\\
\begin{table}[!htb]
\centering
\caption{\label{tab1} The expected number of events N at the future accelerators. The parameters are chosen as $f_{a} = 1$ TeV, $C_{WW} = C_{BB} = 1$, $g_{aee} = 0.64$, the ALP mass $m_{a} = 10$ GeV.  } 
\setlength{\tabcolsep}{2pt}
\begin{tabular}{|c|c|c|c|c|c|c|c|} 
\hline
Collisions& Accelerators& Polarizations& Energy (GeV) & L ($ab^{-1}$) & Run times & $\sigma$ (fb) & N \\
& & & & & (years) &&\\
\hline
\multirow{2}{*}{ $\gamma^{*}e^{-} \rightarrow a e^{-}$  }&\multirow{2}{*}{ ILC} & (0.8, 0.8)&$\sqrt{s} =$ 500 & 4 & 11.5 & 0.3727 & 1490.8\\
 & & &$\sqrt{s} =$1000 & 8 & 13 & 9.2334 & 73867.2\\
\cline{2-8}
  & LHeC & (0.8, 0.8) & $E_{e}$ = 60 & 1 & 6 & 309.046& 309046 \\
\hline
\multirow{2}{*}{$e^{-}e^{+} \rightarrow \gamma a$} &\multirow{2}{*}{ ILC} & (0.8, -0.3)&$\sqrt{s} =$ 500 & 4 & 11.5 & 0.000677 & 2.7\\
 & &(0.8, -0.2) &$\sqrt{s} =$1000 & 8 & 13 & 0.000712 & 5.7\\
\cline{2-8}
 &\multirow{3}{*}{ FCC-ee} &\multirow{3}{*}{ (0, 0)}&$\sqrt{s} =$ 91 & 48 & 4& 45.984 & 2.207 $\times 10^6$\\
 & & &$\sqrt{s} =$240 & 10.8 & 3&0.00113  & 12.2\\
 & & &$\sqrt{s} =$365 & 2.7 & 4&0.00094  & 2.55\\
\hline
\end{tabular} 
\end{table}
\hspace*{0.5cm} Table \ref{tab1} summarizes the expected number of events N for ALP production across proposed future collider facilities, namely the ILC, LHeC, and FCC-ee, evaluated under the parameter set $f_a = 1\text{ TeV}$, $C_{WW} = C_{BB} = 1$, $g_{aee} = 0.64$, and $m_a = 10\text{ GeV}$. The ILC and FCC-ee configurations follow the specifications given in Ref.\cite{blondel}, whereas the LHeC parameters are adopted from Ref.\cite{ahma}. A comparative inspection of the two production topologies reveals distinct energy and luminosity dependencies. For the electron-positron annihilation channel $e^-e^+ \to \gamma a$, the highest discovery potential is achieved at the FCC-ee ($\sqrt{s} = 91\text{ GeV}$) with an integrated luminosity of $\mathcal{L} = 48\text{ ab}^{-1}$, yielding an abundant signal rate of $N \approx 2.207 \times 10^6$ events. Conversely, the photoproduction-like channel $\gamma^* e^- \to a e^-$, the LHeC ($E_e = 60\text{ GeV}$, $\mathcal{L} = 1\text{ ab}^{-1}$) achieves a remarkably large cross-section of $309.046\text{ fb}$, yielding 309046 expected signal events over a 6-year run. 
While this channel at the polarized ILC exhibits substantial signal growth with increasing center-of-mass energy. Under the beam polarization setup $(P_{1}, P_{2}) = (0.8, 0.8)$, the cross-section rises from $0.3727\text{ fb}$ at $\sqrt{s} = 500\text{ GeV}$ ($\mathcal{L} = 4\text{ ab}^{-1}$) to $9.2334\text{ fb}$ at $\sqrt{s} = 1000\text{ GeV}$ ($\mathcal{L} = 8\text{ ab}^{-1}$), corresponding to $1490.8$ and $73867.2$ expected signal events, respectively. Consequently, while the FCC-ee run provides unmatched sensitivity via $e^-e^+ \to \gamma a$, the $\gamma^* e^- \to a e^-$ subprocess emerges as the dominant probe for ALP signatures at sub-TeV and TeV linear colliders.

\section{The decay of ALP 10 GeV}
\hspace*{1cm} In this section, we evaluate the branching ratio of ALP 10 GeV. From Refs.\cite{mbauer1, bieko}, the decay ALP widths to $\gamma\gamma, gg, f\overline{f}$ are expressed by:\\
\begin{equation}
\Gamma(a \rightarrow \gamma\gamma) = \frac{\alpha^{2}m_{a}^{3}}{64\pi^{3}f_{a}^{2}}\left\lvert C_{\gamma\gamma}^{eff}\right\rvert^{2};
\end{equation}
\begin{equation}
\Gamma(a \rightarrow gg) = \frac{\alpha^{2}_{s} m_{a}^{3}}{8\pi^{3}f_{a}^{2}}\left\lvert C_{GG}^{eff}\right\rvert^{2} \left(1 + \frac{83 \alpha_{s}}{4\pi}\right);
\end{equation}
\begin{equation}
\Gamma (a \rightarrow f\overline{f}) = \frac{m_{f}^{2}m_{a}N_{c}^{f}}{8\pi f^{2}_{a}}\left|C_{ff}^{eff}\right|^{2}\sqrt{1-\frac{4m_{f}^{2}}{m_{a}^{2}}}.
\end{equation}

\hspace*{1cm}  Using the parameters chosen as $m_{a} = 10$ GeV, $C^{eff}_{XX}/f_{a} = 1 TeV^{-1}$ \cite{bieko} we evaluate the branching ratio of ALP with mass 10 GeV in Table.\ref{tab2}. The results show that the $b\overline{b}$ channel is dominated when the mass of ALP is 10 GeV. 

\begin{table}[!htb]
\centering
\caption{\label{tab2} Branching ratio of ALP 10 GeV. The parameters are chosen as $m_{a} = 10$ GeV, $C^{eff}_{x}/f_{a} = 1 TeV^{-1}$.} 

\begin{tabular}{|c|c|} 
\hline 
Channels & Br $(\%)$ \\ 
\hline 
$a\rightarrow e^{+}e^{-}$ & $7.103 \times 10^{-7}$ \\
\hline
 $a\rightarrow \mu^{+}\mu^{-}$& 0.0305 \\
\hline
 $a\rightarrow \tau^{+}\tau^{-}$& 8.06\\
 \hline
 $a\rightarrow c \overline{c}$ & 12.781\\
 \hline
  $a\rightarrow b \overline{b}$ & 78.414\\
 \hline
$a\rightarrow \gamma\gamma$ & $1.843 \times 10^{-4}$\\
\hline
$a\rightarrow gg$& 0.714\\
\hline
\end{tabular} 
\end{table}

\hspace*{1cm} Table \ref{tab3} presents the predicted cross-sections for ALP production and subsequent decays into different final states at several proposed future collider facilities. For the process $\gamma^*e^-\to ae^-\to XXe^-$, the results show a substantial enhancement in the cross-sections at the ILC when moving from 500 GeV to 1000 GeV. In particular, the cross-sections increase by approximately a factor of 24.7 for all considered final states, indicating a strong dependence on the collider energy. The LHeC predictions are even larger, exceeding the corresponding ILC results. This indicates the considerable potential of the photon initiated channel at the LHeC, particularly for the hadronic final states. Among the channels considered, the $b\bar b$ final state yields the largest cross-section, reaching 242.335 fb at the LHeC with the energy 60 GeV.  A similar comparison can be made for the process $e^-e^+\to\gamma a\to\gamma XX$ at the ILC. In this case, the cross-sections at 500 GeV and 1000 GeV remain relatively close, with only a modest increase at the higher energy stage. A particularly interesting feature appears at FCC-ee operating at $\sqrt{s}=91$ GeV, where significantly larger cross-sections are obtained for several fermionic final states, especially $b\bar b$, $c\bar c$, $\tau^+\tau^-$, and $\mu^+\mu^-$. In contrast, the corresponding cross-sections at the FCC-ee with 240 GeV and 365 GeV are considerably smaller and exhibit a decreasing tendency with increasing collision energy. These results demonstrate that the expected ALP production rates depend not only on the available center-of-mass energy but also strongly on the collider configuration and the production mechanism. Consequently, different collider facilities may offer complementary opportunities to investigate ALP interactions through their various decay channels.

 \setlength{\tabcolsep}{2pt}
 \begin{longtable}{|c|c|c|c|c|c|c|c|c|}   
 	  	  \caption{ Predicted cross-sections for ALP production followed by its decay into different final states at the proposed future colliders. The upper and lower blocks correspond to the processes $\gamma^*e^-\to ae^-\to XXe^-$ and $e^-e^+\to\gamma a\to\gamma XX$, respectively. Here, $XX$ denotes the final states listed in Table \ref{tab2}, and the beam polarization configurations are specified for each collider setup. }
\label{tab3}\\ 	 
 	   	  \hline
  Accelerators&  Polarizations  &    \multicolumn{7}{|c|}{ $\sigma$ (fb) }\\
 	   	\hline 
\endfirsthead
\multicolumn{9}{c}%
{{\tablename\ \thetable{} -- continued from previous page}} \\
\hline
  Accelerators&  Polarizations  &    \multicolumn{7}{|c|}{ $\sigma$ (fb) }\\
\hline
\endhead

\hline
\multicolumn{9}{r}{{Continued on next page}} \\
\endfoot

\hline
\endlastfoot

 	   	  \multicolumn{9}{|c|}{ $\gamma^{*}e^{-} \rightarrow a e^{-} \rightarrow XX e^{-} $ }\\
 	   	  \cline{1-9}
& & $\sigma_{e^{+}e^{-}e^{-}}$  & $\sigma_{ \mu^{+}\mu^{-}e^{-}}$ &$\sigma_{ \tau^{+}\tau^{-} e^-}$  &   $\sigma_{c \overline{c}e^-}$  & $\sigma_{b \overline{b}e^-}$  & $\sigma_{\gamma\gamma e^-}$  & $\sigma_{gge^-}$\\
\hline
$\text{ILC}_{500} $ & \multirow{2}{*}{(0.8, 0.8)} &2.647 $\times 10^{-9}$ &0.000114 &0.03004 & 0.0476&0.292 &6.869 $\times 10^{-7}$ & 0.00266\\
$\text{ILC}_{1000}$ && 6.558 $\times 10^{-8}$& 0.00282 & 0.744 & 1.180&7.240& 0.0000170&
0.066\\
\hline
$\text{LHeC}_{60}$ & (0.8, 0.8) & 2.195$\times 10^{-6}$& 0.0943& 24.909& 39.499& 242.335& 0.000569& 2.207 \\
\hline
  \multicolumn{9}{|c|}{ $e^{-}e^{+} \rightarrow \gamma a \rightarrow \gamma XX$ }\\
 	   	  \cline{1-9}
& & $\sigma_{\gamma e^{+}e^{-}}$  & $\sigma_{ \gamma\mu^{+}\mu^{-}}$ &$\sigma_{ \gamma\tau^{+}\tau^{-} }$  &   $\sigma_{\gamma c \overline{c}}$  & $\sigma_{\gamma b \overline{b}}$  & $\sigma_{\gamma \gamma\gamma }$  & $\sigma_{\gamma gg}$\\
\hline
 $\text{ILC}_{500} $& (0.8, -0.3) & 4.809$\times 10^{-12}$ & 2.065$\times 10^{-7}$& 0.0000546 &0.0000865&
0.000531 & 1.248$\times 10^{-9}$ & 4.834$\times 10^{-6}$ \\
 $\text{ILC}_{1000}$ & (0.8, -0.2) & 5.057$\times 10^{-12}$  & 2.172$\times 10^{-7}$  & 0.0000574& 0.000091&
0.000558 & 1.312$\times 10^{-9}$ & 5.084$\times 10^{-6}$\\
\hline
$\text{FCC-ee}_{91} $  &\multirow{3}{*}{(0, 0)} & 3.266$\times 10^{-7}$ & 0.014& 3.706& 5.877& 36.058& 0.0000847&0.328 \\
$\text{FCC-ee}_{240} $  && 8.026$\times 10^{-12}$& 3.447$\times 10^{-7}$ & 0.0000911& 0.000144& 0.000886 &
2.083$\times 10^{-9}$ & 8.068$\times 10^{-6}$ \\
$\text{FCC-ee}_{365}$  && 6.677$\times 10^{-12}$& 2.867$\times 10^{-7}$ & 0.0000757& 0.000120& 0.000737 &
1.732$\times 10^{-9}$& 6.712$\times 10^{-6}$ \\
\hline 				 
  \end{longtable}

\newpage
\section{Conclusion}
\hspace*{1cm} ALPs have recently attracted considerable attention as compelling candidates for physics beyond the SM. Their possible connections to fundamental questions in particle physics and cosmology, together with the broad range of theoretically allowed masses and couplings, provide strong motivation for exploring their phenomenological signatures in different experimental environments. In particular, collider searches offer an important complementary approach to astrophysical and cosmological investigations, allowing ALP interactions with SM particles to be examined under controlled and well defined experimental conditions. In this work, future high energy colliders could play a significant role in extending the current exploration of the ALP parameter space.\\
\hspace*{1cm} Motivated by these considerations, we have investigated the prospects for ALP production and detection through two complementary processes, $\gamma^*e^-\to ae^-\to XXe^-$ and $e^-e^+\to\gamma a\to\gamma XX$, considering various fermionic and bosonic final states. Our analysis has focused on the proposed ILC and FCC-ee facilities, as well as LHeC, with particular attention to the effects of collider energy and initial state beam polarization. We have evaluated the dependence of the cross-sections in the ALP production on the scattering angle, the polarization coefficients, the energy and the ALP mass. The numerical results reveal distinct phenomenological features associated with the two production processes. For the photon initiated process ($\gamma^*e^- \to ae^-$), increasing the center-of-mass energy from 500 GeV to 1 TeV at the ILC leads to a substantial enhancement in the predicted cross-sections across the considered final states. In contrast, the associated photon production ($e^-e^+ \to \gamma a$) exhibits a different energy dependence, with a particularly pronounced enhancement at FCC-ee operating at $\sqrt{s}=91$ GeV. The results of cross-section in $e^{-}e^{+} \rightarrow \gamma a$ showed in this work can be compared to the non-resonant ALP production cross section from Belle II searches at the $(4S)$ resonance in detail in Ref.\cite{luzio}. This behavior highlights the importance of the collider energy and production configuration in determining the potential sensitivity to different ALP decay channels. \\


\end{document}